\documentclass[10pt,amsmath,amssymb,aps,pre,letterpaper,twocolumn,floatfix,superscriptaddress]{revtex4-1}
\usepackage{graphicx} 
\usepackage{bm}
\usepackage{textcomp}
\usepackage{gensymb}
\usepackage{array,multirow}
\newcolumntype{K}[1]{>{\centering\arraybackslash}m{#1}}
\usepackage{makecell}
\usepackage{tablefootnote}
\usepackage[english]{babel}
\usepackage{blindtext}

\graphicspath{{./figures/}}

\renewcommand{\(}{\left(}
\renewcommand{\)}{\right)}

\renewcommand\thesection{\arabic{section}}


\begin{document}
	
	\title{Thermal Fracture Kinetics of  Heterogeneous Semiflexible Polymers}
%	\title{Multi-modal transport and dispersion of particles in narrow tubular domains}
	
	\author{Alexander Lorenzo}
	\affiliation{Department of Physics, University of California, San Diego, San Diego, California 92093}
	
	\author{Enrique M. De La Cruz}
	\affiliation{Department of Molecular Biophysics and Biochemistry, Yale University, New Haven, CT 06520}
	
	\author{Elena F. Koslover}
	\email{ekoslover@ucsd.edu}
	\affiliation{Department of Physics, University of California, San Diego, San Diego, California 92093}
%%	\email{smogre@ucsd.edu}
%	\affiliation{Department of Physics, University of California San Diego, La Jolla, California}
%	\author{Elena F. Koslover}
%	\email{ekoslover@ucsd.edu}
%	\affiliation{Department of Physics, University of California San Diego, La Jolla, California}
	\date{\today}
	\preprint{}

%%%%%%%%%%%%%%%%%%%%%%%%%%%%%%%%%%%%%%%%%%%%%%%%%%%%%%%%%%%%%%%%%%%%%%
\begin{abstract}
%%%%%%%%%%%%%%%%%%%%%%%%%%%%%%%%%%%%%%%%%%%%%%%%%%%%%%%%%%%%%%%%%%%%%%

The fracture and severing of polymer chains plays a critical role in the failure of fibrous materials and the regulated turnover of intracellular filaments. 
%While prior studies have focused on the stress-accelerated scission of individual bonds, the role of whole-filament mechanics in thermally driven fragmentation rates has not previously been explored.
% We develop a general model for the fracture of 
 Using continuum wormlike chain models, we investigate the fracture of semiflexible polymers via thermal bending fluctuations, focusing on the role of filament flexibility and dynamics. Our results highlight a previously unappreciated consequence of mechanical heterogeneity in the filament, which enhances the rate of thermal fragmentation particularly in cases where constraints hinder the movement of the chain ends. Although generally applicable to semiflexible chains with regions of different bending stiffness, the model is motivated by a specific biophysical system: the enhanced severing of actin filaments at the boundary between stiff bare regions and mechanically softened regions that are coated with cofilin regulatory proteins. The results presented here point to a potential mechanism for disassembly of polymeric materials in general and cytoskeletal actin networks in particular by the introduction of locally softened chain regions, as occurs with cofilin binding.

\end{abstract}
%\PACS 78.55.Qr \sep 72.20.Ee \sep  05.40.a

\maketitle
\newpage

%%%%%%%%%%%%%%%%%%%%%%%%%%%%%%%%%%%%%%%%%%%%%%%%%%%%%%%%%%%%%%%%%%%%%%
\section{Introduction}
\label{sec:introduction}
%%%%%%%%%%%%%%%%%%%%%%%%%%%%%%%%%%%%%%%%%%%%%%%%%%%%%%%%%%%%%%%%%%%%%%

The fracture properties of polymeric solids pose a key constraint on the manufacture and design of a vast array of man-made materials for load-bearing or weather-resistant purposes\cite{kausch1987polymer,ward2012mechanical}. Furthermore, polymeric materials serve as some of the most important structural and information-bearing components in living organisms, and their rupture (whether through mechanical or environmental stress or through regulated turnover) has a crucial role to play in biological processes ranging from cell division\cite{zhou2015mechanical}, to tumorigenesis\cite{van2001chromosomal}, to cell motility\cite{pollard2003cellular}. Theoretical and experimental explorations of failure mechanisms have established that the fracture of  polymeric solids relies in large part on the scission of individual polymer filaments, with the dynamics and stress-dependence of fracture governed by the kinetics of molecular rupture\cite{regel1972kinetic,tomashevskii1975kinetic,kausch1987polymer}. At the molecular scale, fracture is inherently a thermal process, where the activation energy is lowered by the application of stress on individual bonds along the filament\cite{tomashevskii1975kinetic}.

Fragmentation of a polymer filament is accelerated when externally applied stresses become locally concentrated in specific regions. This principle underlies, for instance, the fragmentation of DNA at discrete folding points under extensional flow\cite{reese1990fracture}, the rupture of microtubules through buckling during spindle reorganization\cite{schiel2011endocytic} and traumatic axonal injury\cite{tang2010mechanical}, and the severing of actin bundles by myosin-driven compression 
 in motile cells\cite{medeiros2006myosin,tsai2019efficient}.
 % and reconstituted contractile networks\cite{murrell2012f}. 
  Local discontinuities in mechanical properties tend to concentrate externally applied stress, leading to preferential fracture of materials at these discontinuous regions\cite{suresh2001graded,cruz2015mechanical}. 

In the case of thermally driven fracture,  the effect of mechanical inhomogeneities in a filament is poorly understood. Prior theoretical work showed that thermal energy is equally partitioned among spatial degrees of freedom in general equilibrium one-dimensional systems\cite{bar2015spatial}. However, fracture is inherently a transient, kinetic process. Understanding fracture rates requires moving beyond equilibrium distributions to 
consider the dynamics of thermal fluctuations in a polymer filament. Here we focus on the role of spatial heterogeneity of mechanical properties in accelerating thermally induced fracture of semiflexible chains.

The general problem of fracture rates in a thermalized, mechanically heterogeneous, polymer filament is motivated in part by a biological system: the cofilin-mediated severing of cytoskeletal actin filaments. Actin is a semiflexible polymer that forms bundles and networks responsible for maintaining cell-scale mechanical properties as well as driving processes such as lamellipodial motility, cytokinesis, and embryonic patterning\cite{fletcher2010cell,salbreux2012actin}. Much of the biological behavior of actin networks relies on the dynamic turnover of individual actin filaments, which is accelerated by the actin-binding protein cofilin. Cofilin assembles cooperatively along actin chains, locally decreasing their bending stiffness and resulting in mechanically heterogeneous partially decorated filaments\cite{mcgough1999ADF,mccullough2008cofilin,pfaendtner2010actin,mccullough2011cofilin,fan2013molecular}. Such filaments fragment, without additional energy input, preferentially at the boundary of cofilinated segments\cite{cruz2009cofilin,mccullough2011cofilin,elam2013biophysics}. While missing bonds at these discontinuities may account for their increased fragility, particularly under stress\cite{schramm2017actin}, an additional contribution to enhanced severing has been proposed that relies
on the concentration of stress at the discontinuities between cofilinated and bare actin segments\cite{cruz2015mechanical,schramm2017actin}. Here, we explore the physical plausibility of enhanced fracture at a junction between soft and stiff regions, in a purely thermal system (ie: in the absence of externally applied compressive forces).

\section{Mechanical model for heterogeneous filament}
\label{sec:model}
%%%%%%%%%%%%%%%%%%%%%%%%%%%%%%%%%%%%%%%%%%%%%%%%%%%%%%%%%%%%%%%%%%%%%%

\begin{figure*}
	\includegraphics[width=\textwidth]{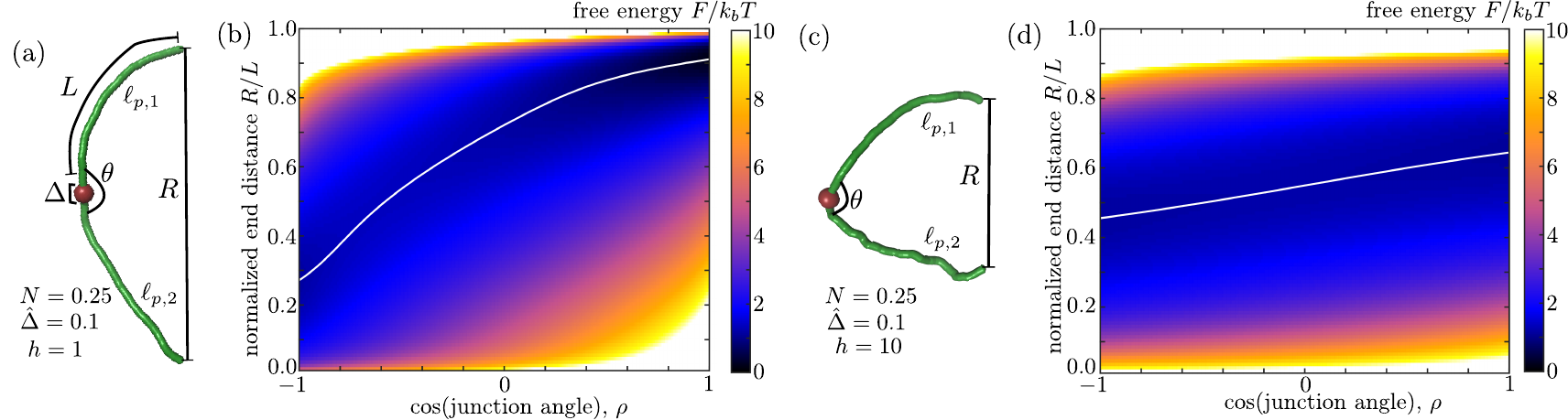}
	\caption{Model schematic and energy landscapes. (a) Sample configuration of homogeneous chain, with model parameters labeled. (b) Free energy landscape for homogeneous chain shown in (a), plotted as a function of junction bending and normalized end distance. White line markes the lowest energy path to steeper junction angles. (c-d) Sample configuration and corresponding free energy landscape for heterogeneous chain with $\ell_{p,1}/\ell_{p,2} =10$. }		
	%	 (a) and heterogeneous (b) wormlike chains. 
	%(b) Sample configuration of heterogeneous chain ($h=10$).
	%	 (c-d) Free energy landscapes for chains with the same parameters as (a) and (b), respectively, plotted as a function of junction bending and normalized end distance. White lines mark the lowest energy path to increasing junction angles. For a homogeneous chain, junction bending requires a greater reduction of the chain end-to-end distance. }
	\label{fig:landscapes}
\end{figure*}

%\subsection{Diblock worm-like chain model formulation}
%\label{subsec:model}
We consider the thermally driven fracture of a mechanically heterogeneous filament, by building upon the well-established continuum ``worm-like chain" (WLC) model for semiflexible polymers\cite{kratky1949wlc,spakowitz2005end}. Prior work on the statistical mechanics of heterogeneous and kinked worm-like chains has established a framework for analytically calculating their distribution functions\cite{wiggins2005exact,zhang2010intrinsic,koslover2013systematic}. Here we focus on the simplest heterogeneous chain: a diblock copolymer consisting of two WLC of equal length $L$ and bending persistence lengths $\ell_{p,1} \geq \ell_{p,2}$. The chains are grafted together at a point junction (Fig.~\ref{fig:landscapes}a), whose bending energy is defined as
\begin{equation}
\begin{split}
\frac{1}{k_b T} E_\text{junc} & = \kappa (1-\rho), \\ \rho & = \cos\theta,
\end{split}
\end{equation}
where $k_bT$ is the thermal energy  and $\theta$ is the bending angle between chain tangents at the junction. The junction represents a short portion of the chain of length $\Delta$, with $\Delta \ll L$. In our model, this junction is treated as a single point with stiffness $\kappa = \ell_{p,1}/\Delta$. In particular, we note that this junction represents the behavior of the last short segment of stiff chain just before the attachment of the softer chain.

%Specifically, we focus on the behavior the last segment of stiff chain just before the attachment of the softer chain. The mechanics of the junction are thus determined by $\kappa = \ell_{p,1}/\Delta$. Our goal will be to explore how the presence of the softer chain just beyond this junction affects the angular dynamics of the junction itself. Because the junction represents a very short region of chain, it is treated as a single point and the slight asymmetry between lengh $L-\Delta$ of the stiff chain and length $L$ of soft chain to either side of it is neglected.

The mechanics of the heterogeneous chain are fully defined by three dimensionless parameters: chain half-length $N = L/(2\ell_{p,1})$, junction length $\hat{\Delta}= \Delta/L = 1/(2\kappa N)$, and heterogeneity $h = \ell_{p,1}/\ell_{p,2}$.

The overall partition function [$G_\text{tot}(\vec{R},\rho)$] for this model is computed from prior results derived for worm-like chains with end constraints\cite{spakowitz2005end,mehraeen2008end,koslover2013systematic}. We start with the partition function [$\hat{G}(\vec{R}, \vec{u}; L,\ell_p)$] for a WLC chain or length $L$, persistence length $\ell_p$, with one end at the origin, the other end at position $\vec{R}$, and final end tangent $\vec{u}$. 
After a Fourier transform from $\vec{R}$ to $\vec{k}$, this function is given by:
\begin{equation}
\begin{split}
\widehat{G}(\vec{k}, \vec{u}; L,\ell_p) = \frac{1}{4\pi}\sum_{l=0}^\infty \mathcal{G}_{0,l}\(2\ell_p k, \frac{L}{2\ell_p}\) P_l(\vec{u}\cdot\vec{k}),
\end{split}
\end{equation}
where $P_l$ are Legendre polynomials and the $\mathcal{G}^m_{l_0,l_f}$ coefficients refer to previously defined continued fraction terms\cite{spakowitz2005end}.
For a heterogeneous wormlike chain with end-to-end vector $\vec{R}$ and junction angle $\theta$, the partition function is obtained from the convolution of two such propagators
\begin{equation}
\begin{split}
G_\text{tot}(\vec{R},\rho; L, \ell_{p,1},\ell_{p,2}) = & \int d\vec{R}_j d\vec{u}_j d\phi \hat{G}(\vec{R_j}, \vec{u_j}; L,\ell_{p,1}) \\
\times & \hat{G}(\vec{R}-\vec{R_j}, \vec{\mathbf{\Omega}(\theta,\phi) \cdot u_j}; L,\ell_{p,2}) \\
\end{split}
\end{equation}
where $\vec{R}_j$ is junction position, $\vec{u}_j$ is the incoming tangent to the junction, and $\Omega(\phi, \theta)$ is a rotation matrix that rotates the cannonical coordinate system by Euler angles $(\phi,\theta, 0)$. The Fourier transform in space, together with an application of the spherical harmonic addition theorem\cite{arfken2005mathematical}, allows this convolution to be simplified to:
\begin{equation}
\begin{split}
\widehat{G}_\text{tot}(\vec{k},\rho; N, 1,h) = \frac{1}{2}\sum_{\ell=0}^\infty P_l(\rho)  \mathcal{G}_{0,l}\(k, N\) \mathcal{G}_{0,l}\(\frac{k}{h}, hN\),
\end{split}
\end{equation}
where we have nondimensionalized all length units by $2\ell_{p,1}$.
Finally, the Fourier inversion is computed according to\cite{mehraeen2008end}
\begin{equation}
\begin{split}
G_\text{tot}(\vec{R},\rho) = \frac{1}{(2\pi^2)(2\ell_{p,1})^3} \int_0^\infty\!\! dk \, \frac{k\sin(2kNr)}{2Nr} \,\widehat{G}_\text{tot}(k,\rho),
\end{split}
\end{equation}
where $r = |R|/(2L)$ is the normalized end separation of the joint chain. This propagator is normalized such that $\int d\vec{R} G_\text{tot}(\vec{R},\rho) = 1$ for each value of $\rho$.

% Namely, the partition function for a fixed end-to-end vector $\vec{R}$ and junction angle (expressed as $\rho$) is given by,
%\begin{widetext}
%\begin{equation}
%\begin{split}
%G_\text{tot}(\vec{R}, \rho) =
%\frac{e^{\kappa \rho}}{4\pi^2  (2\ell_{p,1})^3} 
%\int_0^\infty \! d\hat{k} \; \frac{\hat{k} \sin \left(2kNr\right)}{2Nr} \sum_{l=0}^\infty P_l(\rho)\mathcal{G}_{0,l}^0\(k,N\) 
%\mathcal{G}_{l,0}^0\(\frac{k}{h},hN\),
%\end{split}
%\end{equation}
%\end{widetext}
%where $r = |\vec{R}|/(2L)$ is the normalized end separation and $\mathcal{G}^{0}_{l_0,l_f}$ refer to previously defined continued fraction terms\cite{mehraeen2008end}. 

The free energy ($F$) of the chain is then defined  as the log of the partition function, with an additional term for the bending of the junction angle. Namely,
\begin{equation}
\begin{split}
\frac{1}{k_b T}F(r,\rho) = \kappa(1-\rho) - \log \left[ r^2 G_\text{tot}(r,\rho)\right]
\end{split}
\label{eqn:F}
\end{equation}
 This free energy landscape is plotted in Fig.~\ref{fig:landscapes} for a homogeneous, stiff chain and a heterogeneous chain.
%($L = 0.5 \ell_{p,1}$) and a heterogeneous chain.
%ly stiff chain and a heterogeneous chain. 

We focus on filament fracture at the junction point, assuming that fracture will occur when thermal fluctuations push the junction energy ($E_\text{junc}$) above some predefined cutoff ($E^*$). This model represents a fracture process where the junction must hop over a transition energy barrier, with the cosine of the bending angle $\rho$ as the reaction coordinate. Chains with a more flexible junction (lower $\kappa$)
will have to reach more extreme junction bending ($\rho^* = 1-E^*/\kappa$) than chains with a more stiff junction (higher $\kappa$). 
This model is consistent with previous analyses of experimental data on fracture of short cofilin-decorated actin filaments that points to fracture occuring beyond a critical bending angle  that increases with lower filament persistence length\cite{mccullough2011cofilin}. Critical energies of approximately $22$kT have been estimated for the severing of bare actin filaments\cite{mccullough2011cofilin}.

The overall rate of fracture is obtained from the mean first passage time (MFPT) to the critical value $\rho^*$, as the system fluctuates thermally over the free energy landscape plotted in Fig.~\ref{fig:landscapes}.
The kinetics of fracture are thus determined by a free energy barrier incorporating both the junction bending energy and the configurational free energy of the worm-like chains. For a homogeneously stiff chain, surmounting this barrier along the minimum energy path requires bringing the ends of the chain closer together (Fig.~\ref{fig:landscapes}c). For the heterogeneous chain, by contrast, the cutoff junction angle can be reached without substantial change in the end-to-end distance (Fig.~\ref{fig:landscapes}d). The importance of this effect in determining the overall time to fracture depends on the dynamics of the end-to-end coordinate $r$ compared to the dynamics of the junction angle.

%%%%%%%%%%%%%%%%%%%%%%%%%%%%%%%%%%%%%%%%%%%%%%%%%%%%%%%%%%%%%%%%%%%%%%
\section{Dynamics over Free Energy Landscape}
\label{sec:kinetics}
%%%%%%%%%%%%%%%%%%%%%%%%%%%%%%%%%%%%%%%%%%%%%%%%%%%%%%%%%%%%%%%%%%%%%%

\begin{figure*}
	\includegraphics[width=\textwidth]{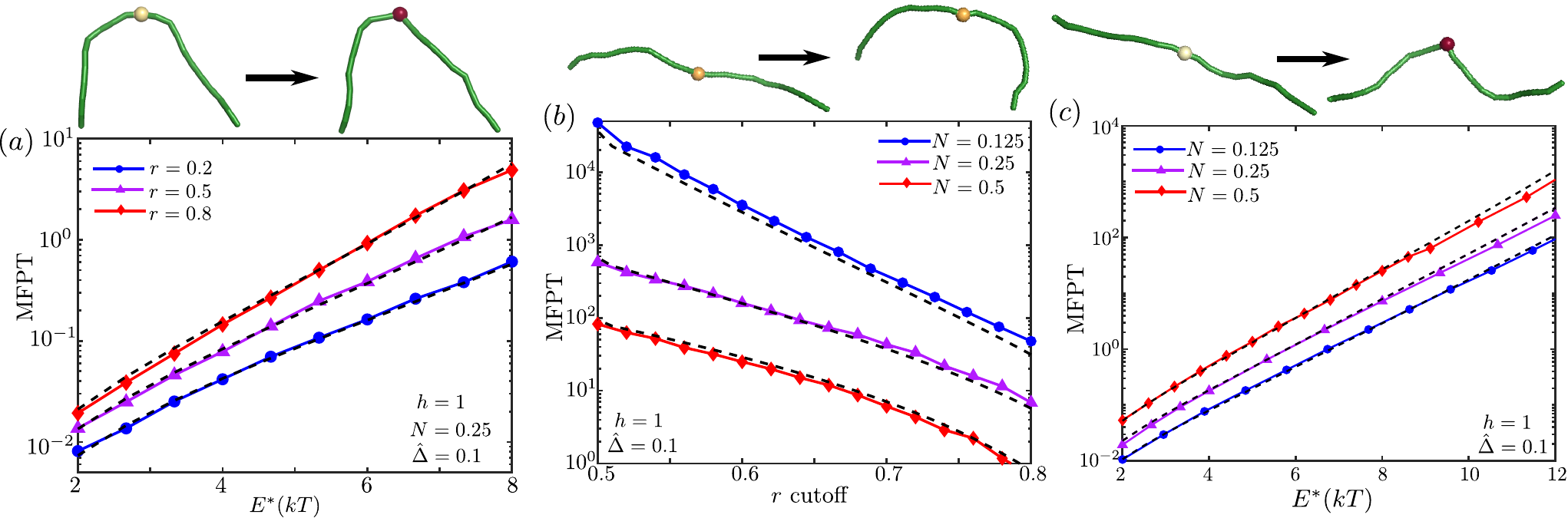}
	\caption{Comparison of approximate dynamics over free energy landscapes versus Brownian dynamics simulations. (a) MFPT to a cutoff junction energy $E^*$, for fixed end distance. (b) MFPT to a cutoff value of the normalized end-to-end distance $r$. (c)
		MFPT to a junction energy $E^*$, with free chain ends. In all cases, dashed black lines correspond to first passage times calculated from the free energy landscapes, solid lines correspond to Brownian dynamics simulations. All times are non-dimensionalized by $D^{(0)}_\rho$.% Homogeneous chains with $h=1$ are used for these simulations.
		Top panels show example start and end configurations, with junction color indicating energy at the junction. }		
	\label{fig:bdsim}
\end{figure*}
%We use a simplified model for kinetics over an energy landscape to estimate the rates of fracture, validating the model and the appropriate choice of dynamic prefactors by comparison to Brownian dynamics simulations (Fig.~\ref{bdsim}; details in Supplemental Information (SI)\cite{supplemental}).
%appropriate prefactors for dynamics of the $\rho$ and $r$ coordinates by comparison to Brownian 

To calculate kinetics over the free energy landscape, we make the simplifying assumption that for each value of the end distance $r$, the kinetics of transition to the cutoff $\rho^*$ can be described by a single time-scale --- the mean first passage time $\tau(\rho^*; r)$ along a horizontal slice of the landscape. Dynamics along the angular coordinate are defined by a variable friction coefficient that depends on the value of the junction angle,
\begin{equation}
\begin{split}
\zeta(\rho) = \frac{k_BT}{D^{(0)}_\rho} \frac{5-3\rho}{6(1-\rho^2)}
\end{split}
\label{eqn:fric}
\end{equation}
where $D^{(0)}_\rho = \frac{k_BT}{\mu \Delta^3}$ and $\mu$ is the translational friction coefficient per unit length of the chain. This expression is derived from the dynamics of two connected rigid links (Appendix \ref{app:rhodynamics}). We note that the numerical prefactor in the friction coefficient (Eq.~\ref{eqn:fric}) depends on  mapping from the behavior of two rigid links to the dynamics of a point-like kink representing a short length of semiflexible chain. This prefactor is not directly determined by our theory, and we fit the appropriate length of links to establish this mapping (specifically, $\ell_\text{link} = 2\Delta$) by matching the theory to Brownian dynamics simulations of semiflexible chains, as described below. This is the only fitting parameter in the model, a single value (as shown in Eq.~\ref{eqn:fric}) is used for all results described.

%~\cite{supplemental}. %[see Supplemental Information (SI)~\cite{supplemental}].
% [MOVIES! INCLUDING FAN-LIKE PIC ZOOMED INTO JUNCTION]].
  The mean first passage time over a one-dimensional landscape can be computed from the Fokker-Planck  equation\cite{fox1986functional,risken1996fokker,koslover2009twist}, appropriately modified for spatially varying diffusivity\cite{lau2007state} (Appendix \ref{app:kolmogorov}). Specifically, the fracture time for a fixed value of $r$ is given by
  \begin{equation}
  \begin{split}
&    \tau(\rho^*,r) = \frac{1}{\int_{\rho^*}^1 e^{-F(r,\rho)/k_bT} d\rho} \times \\
   &\times \left[
   \int_{\rho^*}^1 d\rho \int_{\rho^*}^\rho d\rho' \int_{\rho'}^1 d\rho'' \frac{\zeta(\rho')}{k_bT} e^{\(F(r,\rho')-F(r,\rho'')-F(\rho)\)/k_bT}
   \right].
  \end{split}
  \end{equation}
  
   %to allow for the correct long-time equilibrium behavior. 
   Brownian dynamics simulations of a discretized WLC model are used to validate our calculations of the mean first passage time %to junction energy $E^*$
    for fixed values of the end distance, with the free energy landscape given by Eq.~\ref{eqn:F} and the friction coefficient in Eq.~\ref{eqn:fric}. As shown in Fig.~\ref{fig:bdsim}a, the continuum chain theory accurately reproduces the kinetics of reaching a high junction angle in simulations.
    
%     (Fig.~\ref{fig:bdsim}a)\cite{supplemental}.
    % We note that the numerical prefactor ($1/6$) in the friction coefficient for the angular coordinate (Eq.~\ref{eqn:fric}) depends on the mapping from two rigid links to the dynamics of a point-like kink representing a short length 
    %; details in SI\cite{supplemental}).

The overall mean first passage time to fracture can be computed by considering a system that fluctuates over discrete states in the normalized end distance, with state $i$ corresponding to  $r_i = i \delta r$, and the discretization set to $\delta r = 0.01$.
%$(i-1) \delta r < r < i \delta r$. 
%We use a discretization of $\delta r = 0.01$. 
The system is assumed to start in thermal equilibrium, with the probability of starting in state $i$ set by a Boltzmann factor corresponding to the free energy of that state:
% $F_i = -\log r^2 \int d\rho \exp[- G_\text{tot}(r,\rho)]$
$F_i = - k_BT \log \int d\rho \exp[- F(r_i,\rho)/k_BT]$.
 Transitions between states occur with rate constants $k_i^{(\pm)}$, given by
 \begin{equation}
 \begin{split}
 k_i^{(\pm)} & = \frac{\frac{k_R}{\delta r^2} (F_{i\pm 1} - F_i)}{\exp(F_{i\pm 1} - F_i) - 1}, \\
 \end{split}
 \label{eqn:ki}
 \end{equation}
as derived from a discretization of the Fokker-Planck equation\cite{wang2003robust}. The dynamic prefactor is taken to be  the time-scale for three-dimensional translational diffusion of a chain of length $L$ over a length scale $\Delta R = 2L\delta r$, according to:
\begin{equation}
\begin{split}
 \frac{k_R}{\delta r^2} & = \frac{6k_BT}{(\mu L) (2L\delta r)^2}.
\end{split}
\label{eqn:kr}
\end{equation}

This approximate model for dynamics in the end-to-end distance 
gives comparable results to Brownian dynamics simulations that sample the average time required to reach a cutoff end-to-end distance for chains starting in thermal equilibrium  (Fig.~\ref{fig:bdsim}b). 
These simulations are compared against an analytical model for dynamics over a discrete one-dimensional landscape\cite{koslover2012force}, with the transition rates given by Eq.~\ref{eqn:ki}, \ref{eqn:kr}. 

%Comparison to Brownian dynamics simulations (Fig.~\ref{fig:bdsim}b,c) indicates that a good estimate for the chain end dynamic prefactor is given by $k_R/(\delta r)^2 = \frac{6k_BT}{(\mu L) (2L\delta r)^2}$, corresponding to three-dimensional translational diffusion of a chain of length $L$, over a length scale $2L\delta r$. The dimensionless parameter describing the rate of equilibration in chain end distance compared to the junction angle is then %$k_R/D_\rho^{(0)} = \frac{3}{16 (\delta r)^2} \(\frac{1}{\kappa N}\)^3$.
%$k_R/D_\rho^{(0)} = \frac{3}{2} \widehat{\Delta}^3$. 

To put together both dimensions of the energy landscape, we treat fracture as a Poissonian process within each particular end-distance state, with average time given by $\tau_i = \tau(\rho^*, r_i)$. 
%within each end-distance state fracture is treated as a Poissonian process with average time $\tau_i = \tau(\rho^*, r_i)$. 
We compute the overall mean time to fracture for a system that fluctuates over these states using a matrix inversion method, as described in previous work on the kinetics of systems with fluctuating rates\cite{koslover2012force}. This approach for representing the dynamics of the system as movement over a two-dimensional free energy landscape is validated by comparison to Brownian dynamics simulations with unconstrained homogeneous chains (Appendix \ref{app:bdsim}; Supplemental Videos 1, 2). As shown in Fig.~\ref{fig:bdsim}c, Our model of dynamic fluctuations over a two-dimensional energy landscape thus appears to accurately represent the behavior of simulated chains, with respect to the time required for a junction region to hit a cutoff angle.

%\section{Fracture of Heterogeneous Chains with Constrained Ends}
\section{Fracture rates for heterogeneous chains}

\begin{figure*}
	\includegraphics[width=\textwidth]{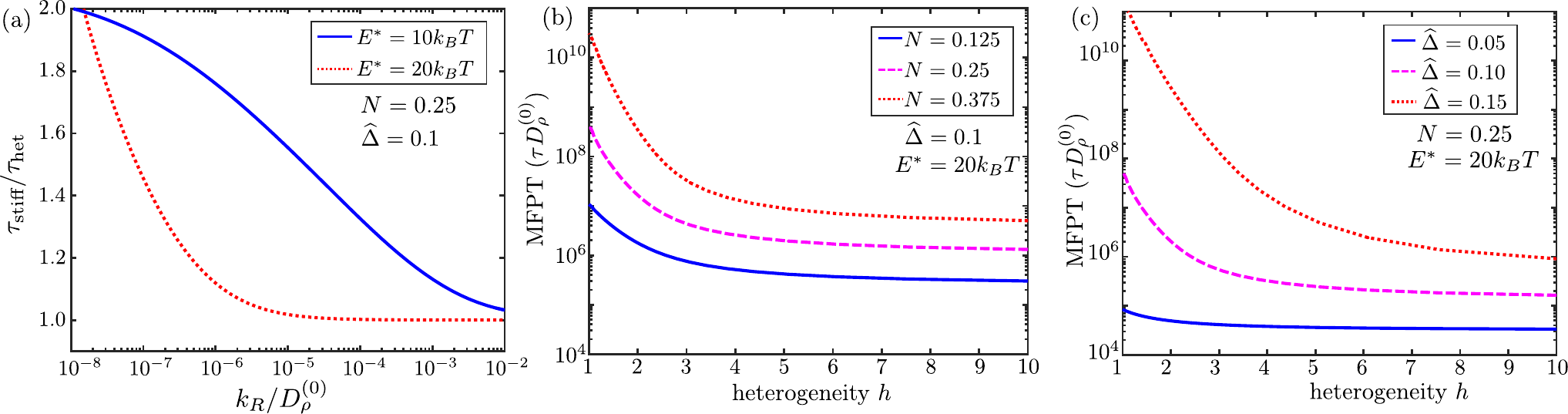}
	\caption{Chain heterogeneity enhances junction fracture rates when chain end dynamics are slow.
		(a) Ratio of MFPT to fracture for uniformly stiff ($h=1$) and heterogeneous ($h=10$) chains is plotted versus the relative rate of chain end dynamics compared to junction dynamics. (b-c) Time to fragmentation in the limit of infinitely slow chain end dynamics. Dimensionless MFPT
		%to a cutoff junction energy $E^*=20k_BT$
		is shown as a function of heterogeneity $h$ for (b) chains with a fixed junction length $\widehat{\Delta} = 0.1$ and varying stiffness and (c) chains with a fixed stiffness ($N=0.25$) but varying junction length. Chains are assumed to start from an equilibrium distribution.}
	\label{fig:dynamics}
\end{figure*}

%[CITE 10-40KT FROM MCCULLOUGH 2011 PAPER, SCHRAMM PAPER FOR 40 KT] 

%We assume that the chain begins in a configuration selected from its equilibrium distribution and consider the mean time to fracture for two limiting cases of the end-to-end dynamics (Fig.~\ref{fig:dynamics}a). 
The overall time to fracture is dependent on the relative rate of motion in the end-to-end distance as compared to the rate of junction fluctuations (Fig.~\ref{fig:dynamics}a). 
For the case of very rapid end equilibration (high $k_R/D_\rho^{(0)}$), the chain would be expected to sample all end positions over a time-scale that is short compared to the fracture time. 
%The overall rate of fracture is then obtained as the average of the rate constants $1/\tau_i$, weighted by the equilibrium probability of starting at each state.  
 In this limit, the fracture dynamics are determined entirely by the stiffness and friction coefficient for the junction bending ($\rho$) and are independent of the mechanical properties of the rest of the chain. 
 
The opposite regime holds when the dynamics of the end distance are much slower than those of the junction angle. In this case, the end distance remains constant at its starting value, and the mean time to fracture is the weighted average of the individual $\tau_i$. 
%The distinction between taking the average of kinetic rates in the case of rapid equilibration versus the average of first passage times in the case of slow equilibration was previously discusses in the general context of kinetics determined by rapidly fluctuating forces\cite{koslover2012}. 
Softer mechanics in one half of the chain make it more probable that a lower value of $r$ will initially be selected from the equilibrium distribution. This lower $r$ persists over time and allows the junction to more rapidly reach the cutoff angle. 

%\begin{figure}
%	\includegraphics[width=8.6cm]{slowRfig}
%	\caption{Time to fragmentation in the limit of infinitely slow chain end dynamics. Dimensionless mean first passage time to a cutoff junction energy $E^*=20k_BT$ is shown as a function of heterogeneity $h$ for (a) chains with a fixed junction length $\widehat{\Delta} = 0.1$ and varying stiffness and (b) chains with a fixed stiffness ($N=0.25$) and varying junction length. Chains are assumed to start from an equilibrium distribution with end-to-end distance kept fixed over time. }
%	\label{fig:slowR}
%\end{figure}

Fig.~\ref{fig:dynamics}b,c show the mean time to fracture for chains with different degrees of heterogeneity $h$ in the case of fixed end-to-end distance (infinitely slow $r$ dynamics). In this limit, a purely stiff chain will be slow to reach fracture at the junction because a higher overall chain deformation energy is required to bend the junction to the point of fracture. A purely soft chain will also be slow to reach fracture because the requisite junction angle $\theta^*$ to achieve the same cutoff energy will be correspondingly larger\cite{mccullough2011cofilin}. Rapid fracture can be achieved by a heterogeneous chain, where the junction stiffness and hence the cutoff angle are set by the stiff side of the chain, while the low persistence length of the soft side enables the junction to reach that cutoff angle without moving the chain ends or incurring a substantial cost in chain deformation energy. The enhancement due to chain heterogeneity can reach several orders of magnitude in cases where the junction must reach very steep bending angles in order to  fracture (high $N$ and $\widehat{\Delta}$).

%While our model focuses on fracture at a specific junction region, at the end of the stiff chain segment, it provides insight into the behavior of other regions in a heterogeneous chain. In particular, regions on the soft side of the chain will behave similarly to the center of a homogeneously soft chain, and are expected to be slow to fracture because of the steep bending angle required to reach the cutoff energy. By contrast, regions located in the middle of the stiff side would require moving large segments of stiff chain to assume the requisite bending angle, again leading to slow fracture dynamics. Rapid dynamics become possible specifically close to the junction between the soft and stiff regions

%We note that the model described here differs from previous athermal models for fracture\cite{cruz2015mechanical} which indicated that a heterogeneous chain concentrates stresses at the junction point when the chain is forced into a buckled configuration. The effect we describe here occurs despite the fact that the initial configuration of the chain is allowed to sample from the equilibrium distribution. In fact, the contrast between the case of rapid and slow $r$ equilibration highlights that the ability for heterogeneous chains to accelerate thermal fracture is a dynamic rather than an equilibrium effect. 

%\section{Fracture of Heterogeneous Chains with Free Ends}
In the case where chain ends are unconstrained, calculating the fracture rate requires an estimation of the rate of chain end dynamics compared to the dynamics in the junction coordinate. 
%Comparison to Brownian dynamics simulations (Fig.~\ref{fig:bdsim}b,c) indicates that a good estimate for the chain end dynamic prefactor is given by $k_R/(\delta r)^2 = \frac{6k_BT}{(\mu L) (2L\delta r)^2}$, corresponding to three-dimensional translational diffusion of a chain of length $L$, over a length scale $2L\delta r$. The dimensionless parameter describing the rate of equilibration in chain end distance compared to the junction angle is then %$k_R/D_\rho^{(0)} = \frac{3}{16 (\delta r)^2} \(\frac{1}{\kappa N}\)^3$.
The dimensionless parameter describing these relative rates is
$k_R/D_\rho^{(0)} = \frac{3}{2} \widehat{\Delta}^3$. 

\begin{figure}
	\centerline{\includegraphics[width=7.6cm]{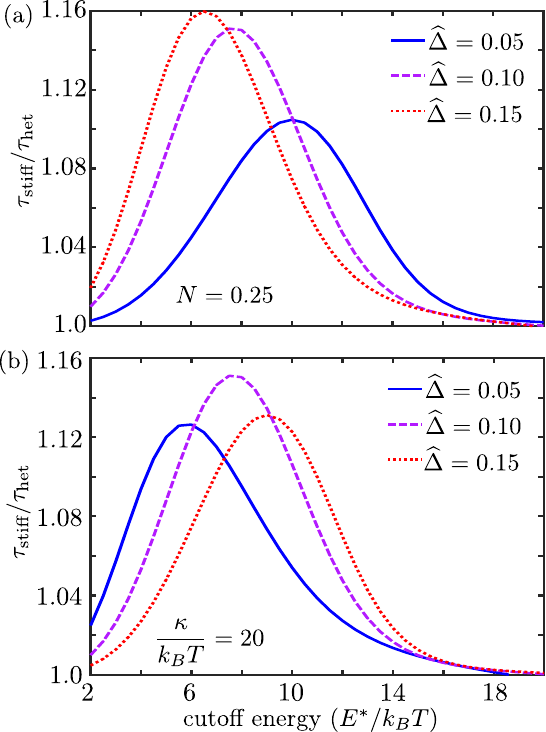}}
	\caption{Enhacement in fracture rate for a heterogeneous vs homogeneous chain with free chain ends. The ratio of MFPT to fracture for a fully stiff chain ($h=1$) vs a heterogeneous chain ($h=10$) is plotted as a function of the cutoff energy. (a) Filaments with constant length $N$ but varying junction size and stiffness (solid: $\kappa = 40$, dashed: $\kappa=20$, dotted: $\kappa=13$. (b) Filaments with varying length but constant junction stiffness $\kappa$ (solid: $N=0.125$, dashed: $N=0.25$, dotted: $N=0.5$). }
	%	\caption{Enhacement in fracture rate for a heterogeneous vs homogeneous chain with free chain ends. The ratio of MFPT to fracture for a fully stiff chain ($h=1$) vs a heterogeneous chain ($h=10$), computed using approximate dynamics over a free energy landscape, is plotted as a function of the cutoff energy, for different dimensionless chain lengths $N$. The junction size is taken to be $\hat{\Delta}$ in all cases.}
	\label{fig:freeends}
\end{figure}

For chain heterogeneity to enhance thermal fracture, this ratio of rates must be small (ie: the sampling of junction angles must be substantially faster than the end-to-end motion).
%THE SAMPLING OF JUNCTION ANGLES MUST BE SUBSTANTIALLY FASTER THAN THE END-TO-END MOTION.
%the junction dynamics must be substantially faster than the end-to-end motion. 
However, when $\Delta$ becomes small for a chain of constant length, the junction stiffness $\kappa$ must increase and the fracture process becomes dominated by junction energetics rather than deformation of larger portions of the chain.
 %necessarily increases. The cutoff angle which must be reached for fracture is then decreased and the fracture process becomes dominated by the energetics of the junction rather than the deformation of larger portions of the chain.
  %As seen in Fig.~\ref{fig:dynamics}c and \ref{fig:freeends}a, 
  In this limit the fracture rate becomes similar for heterogeneous and homogeneously stiff chains (Fig.~\ref{fig:dynamics}c, \ref{fig:freeends}a). If the end-to-end dynamics are slowed down by increasing the chain length $L$ while keeping the junction length $\Delta$ constant (Fig.~\ref{fig:freeends}b), then the stiff side of the chain becomes more flexible and the fracture enhancement from chain heterogeneity decreases. Overall, the enhancement in fracture rates for a heterogeneous chain with free end conditions maxes out at approximately $15\%$, for a heterogeneity $h = \ell_{p,1}/\ell_{p,2} = 10$ (Fig.~\ref{fig:freeends}). 
  %It should be noted that as $\hat{\Delta}$ increases, the size scale associated with the kink becomes comparable to the size of the filament itself, and the approximation of the kink as a point junction breaks down.

%%%%%%%%%%%%%%%%%%%%%%%%%%%%%%%%%%%%%%%%%%%%%%%%%%%%%%%%%%%%%%%%%%%%%%
\section{Concluding Remarks}
\label{sec:conclusion}
%%%%%%%%%%%%%%%%%%%%%%%%%%%%%%%%%%%%%%%%%%%%%%%%%%%%%%%%%%%%%%%%%%%%%%

Our calculations show that filament heterogeneity can substantially enhance the rate of thermal fracture in the case of restricted end-to-end dynamics of the filament. A modest enhancement is expected for the case of a chain with freely moving ends. We note that the model developed here differs from previous athermal models for fracture\cite{cruz2015mechanical, schramm2017actin} which indicated that a heterogeneous chain concentrates stresses at the junction  when the chain is forced into a buckled configuration. The enhancement in thermally driven fracture  occurs despite the fact that the initial configuration of the chain is allowed to sample from the equilibrium distribution. %It is a purely dynamic effect arising from the  
The contrast between the case of rapid and slow $r$ equilibration (Fig.~\ref{fig:dynamics}a) highlights the purely dynamic nature of this effect. Fracture enhancement arises from the separation in timescales between fluctuations at the junction  versus moving the ends of the entire polymer. The presence of a softer chain region allows a junction to reach steep bending angles without requiring large movements of the chain ends and without paying a large energetic cost for the chain deformation. %For the case with free ends, this is a purely dynamic effect, stemming from the slower timescales for equilibrating 

While our model focuses on fracture in a specific junction region, at the internal end of the stiff chain segment, it provides insight into the behavior of other regions in a heterogeneous chain. In particular, regions on the soft side of the chain will behave similarly to the center of a homogeneously soft chain, and are expected to be slow to fracture because of the steep bending angle required to reach the cutoff energy. By contrast, regions located in the middle of the stiff side would require moving large segments of stiff chain to assume the requisite bending angle, again leading to slow fracture dynamics. Rapid dynamics become possible specifically close to the junction between the soft and stiff regions, but where the junction itself is still stiff.

The model with restricted chain ends is particularly relevant for the cofilin-mediated severing of actin filaments within a cytoskeletal network. In such networks cross-links and entanglements can effectively restrict the movement of certain positions along the chain, while allowing rapid equilibration of chain positions between the cross-link points. Our results indicate that in such situations introducing mechanical heterogeneity into the actin filaments by cofilin binding  should substantially enhance thermal severing rates. 

It should be noted that, in addition to changing the flexibility of actin filaments, cofilin binding also alters the filament twist density. Recent experiments have shown that constraining filaments to prevent torsional equilibration enhances actin filament severing by cofilin
%[PAVLOV AND REISLER PAPER, 2007, ACTIN FILAMENT SEVERING BY COFILIN; ALSO ELAM REVIEW], 
%both in the case of cross-linked networks and externally induced end constraints
\cite{pavlov2007actin,wioland2019torsional,elam2013biophysics}. The effect described here centers on severing due to bending fluctuations and may provide a parallel, unrelated mechanism for cofilin-driven fracture. Both twist-based and bending-based severing are expected to depend on the density and mechanics of cross-links in an actin network. By providing a feedback mechanism between network structure and actin severing dynamics, these physical effects may play an important role in regulating the self-assembly, turnover, and mechanoresponse of cytoskeletal structures.

In addition to helping unravel the mechanisms of actin severing by cofilin, the results presented here are generally applicable to the fracture of any semiflexible thermally fluctuating polymer. Enhanced rates of thermally-activated fracture in mechanically heterogeneous chains point towards general principles for controlling the stability of nanoscale systems, including polymer networks, nanotubules, and molecular threads, for a broad range of biological and industrial applications.

%%%%%%%%%%%%%%%%%%%%%%
\begin{acknowledgements}
	The authors thank R. Phillips, H. Garcia, J. Kondev, and J. Theriot for organizing the Physical Biology of the Cell course, whence this collaboration originated. 
	%[ROB/JANE/JULIE FOR BRINGING US TOGETHER AT PBOC]. 
	Funding was provided by the Alfred P. Sloan Foundation Fellowship (E.F.K.), the William A. Lee Undergraduate Research Award (A.L.) and NIH grant R01-GM097348 (E.D.L.C). 
	%E.C. SUPPORTED THROUGH NIH GRANT R01-GM097348.
\end{acknowledgements}

\renewcommand{\theequation}{\Alph{section}-\arabic{equation}}
% redefine the command that creates the equation no.
\setcounter{equation}{0}  % reset counter 
\setcounter{subsection}{0}  % reset counter 
\renewcommand{\thesection}{\Alph{section}}
\vspace{1cm}
\centerline{\large\bf APPENDIX}%\vspace{-0.2\baselineskip}
\setcounter{section}{0}
%\addtocounter{section}{1}

%\renewcommand{\thefigure}{S\arabic{figure}}
%\setcounter{figure}{0}
\renewcommand{\theequation}{S\arabic{equation}}
\setcounter{equation}{0}

%%%%%%%%%%%%%%%%%%%%%%%%%%%%%%%%%%%%%%%%%%%%%%%%%%%%%%%%%%%%%%%%%%%%%%
\section{Angular dynamics for coupled rigid links}
\label{app:rhodynamics}
%%%%%%%%%%%%%%%%%%%%%%%%%%%%%%%%%%%%%%%%%%%%%%%%%%%%%%%%%%%%%%%%%%%%%%

In this section we derive the angular dynamics for two connected rigid links, each of length $\ell$, in a highly viscous fluid. We assume each of the links has a friction coefficient per unit length $\mu$, and that there is a bending modulus $\kappa$ for the junction between the links. This simplified system serves as a basis for deriving the appropriate dynamics of the junction angle for the heterogeneous worm-like chain. 

We define a given configuration of the system by the center of mass positions for the two rigid rods ($\vec{r}_1, \vec{r}_2$) and their normalized orientations ($\vec{u}_1, \vec{u}_2$). The overall energy for this configuration is then given by,
\begin{equation}
\begin{split}
E = \kappa (1 - \vec{u}_1\cdot\vec{u}_2) + \vec{\lambda} \cdot \(\vec{r}_1 + \frac{\ell}{2} u_1 - \vec{r}_2 + \frac{\ell}{2}\vec{u}_2\).
\end{split}
\end{equation}
Here, the first term corresponds to the bending energy of the junction between the two rods and the second term uses a Lagrange multiplier ($\vec{\lambda}$) to enforce the connectivity of the two inextensible rods at the junction.

In the freely draining approximation, and in the absence of Brownian forces, the overdamped dynamics of such a system are defined by the equations,
\begin{equation}
\begin{split}
\zeta_r \vec{\omega}_i & = - \vec{u}_i \times \frac{\partial E}{\partial u_i}\\
\zeta_t \frac{d\vec{r}_i}{dt} & = - \frac{\partial E}{\partial \vec{r}_i},
\end{split}
\end{equation}
where $\vec{\omega}_i$ gives the rotational velocity for each rod ($i=1,2$). Here, $\zeta_r = \mu \ell^3/12$ is the rotational frictional coefficient of each rod around its center of mass and $\zeta_t = \mu \ell$ the translational friction coefficient\cite{doi1988theory}. The Lagrange multiplier $\vec{\lambda}$ can be obtained from the constraints:
\begin{equation}
\begin{split}
\frac{d}{dt} \(\vec{r}_1 + \frac{\ell}{2} u_1 - \vec{r}_2 + \frac{\ell}{2}\vec{u}_2\) \cdot \vec{u}_1 = 0 \\
\frac{d}{dt} \(\vec{r}_1 + \frac{\ell}{2} u_1 - \vec{r}_2 + \frac{\ell}{2}\vec{u}_2\) \cdot \vec{u}_2 = 0. \\
\end{split}
\end{equation}
Solving these equations yields $\vec{\lambda}\cdot\vec{u}_i = \frac{6\kappa(1-\rho^2)}{\ell (5 - 3\rho)}$, where $\rho = \vec{u}_1\cdot \vec{u}_2$. The dynamics of the angular coordinate $\rho$ are then given by,
\begin{equation}
\begin{split}
\frac{d\rho}{dt} = \frac{48 \kappa (1-\rho^2)}{\mu \ell^3 (5-3\rho)}.
\end{split}
\end{equation}

This expression gives the effective friction coefficient for the coordinate $\rho$ according to
\begin{equation*}
\begin{split}
\frac{d\rho}{dt} & = -\frac{1}{\zeta(\rho)} \frac{\partial E_\text{bend}}{\partial \rho} =  -\frac{\kappa}{\zeta(\rho)} \\
\zeta(\rho) & = \frac{\mu \ell^3 (5-3\rho)}{48 (1-\rho^2)}
\end{split}
\end{equation*}

For the angular dynamics of a junction in a continuum worm-like chain, changes in the angle require dragging along a length of chain that should scale as the junction size $\Delta$. We select an effective link length $\ell=2\Delta$ for use in Eq.~\ref{eqn:fric}. The prefactor of $2$ is obtained from fitting to Brownian dynamics simulations with fixed end-to-end distance (as shown in Fig.~\ref{fig:bdsim}a). This is the only fitting parameter in the theory and is used for all results shown in subsequent figures.
%By comparing Brownian dynamics simulations with calculations of first passage times on a free energy landscape over the angular coordinate (Fig.~\ref{fig:bdsim}a), we find that setting $\ell=2\Delta$ gives an accurate representation of the dynamics.

%%%%%%%%%%%%%%%%%%%%%%%
\section{Mean first passage time on a 1D landscape}
\label{app:kolmogorov}
%%%%%%%%%%%%%%%%%%%%%%%%%
For one-dimensional systems with spatially varying diffusivity $D(x)$ and free energy landscape $F(x)$, it has been shown that the Fokker-Planck equation which correctly reproduces the Boltzman distribution in the steady state\cite{lau2007state} is given by,
\begin{equation}
\begin{split}
\frac{dG(x,t|x_0)}{dt} = \frac{\partial}{\partial x} \left[D(x) \(\frac{1}{kT}\frac{\partial F}{\partial x}G + \frac{\partial G}{\partial x}\)\right]
\end{split}
\end{equation}
where $G(x,t|x_0)$ is the Green's function giving the distribution over $x$ at time $t$ for a system that started at position $x_0$. A corresponding backward Kolmogorov equation can be derived for this system\cite{risken1996fokker} as,
\begin{equation}
\begin{split}
\frac{dG}{dt} = \left[-\frac{D(x_0)}{kT}\frac{\partial F}{\partial x_0} + \frac{\partial D}{\partial x_0}\right] \frac{\partial G}{\partial x_0} + D(x_0) \frac{\partial^2 G}{\partial x_0^2}
\end{split}
\label{eqn:reversekolmogorov}
\end{equation}

Assuming the system has an absorbing boundary at $a$ and a reflecting boundary at $L$, the
mean first passage time is defined based on the probability $Q(t|x_0) = \int_a^L G(x,t|x_0) dx$ that the absorbing boundary has not yet been reached. Namely, the MFPT is given by $T(x_0) = -\int_0^\infty t \frac{dQ}{dt}$. We solve for $T(x_0)$ using Eq.\ref{eqn:reversekolmogorov} in a manner analogous to previous calculations with a constant diffusivity\cite{fox1986functional,koslover2009twist}. Assuming an equilibrated distribution of starting positions, the overall mean first passage time is then given by
\begin{equation}
\begin{split}
& \left<T\right> = \frac{1}{\int_a^L e^{-F(x)/kT} dx} \times \\
&\times \left[
\int_a^L dx \int_a^x dy \int_y^L dz \frac{1}{D(y)} e^{\(F(y)-F(z)-F(x)\)/kT}
\right]
\end{split}
\label{eqn:mfpt}
\end{equation}

We use numerical integration of Eq.~\ref{eqn:mfpt} to calculate the mean first passage time for each fixed value of $r$ over the energy landscape plotted in Fig.~\ref{fig:landscapes}.

\bigskip
%%%%%%%%%%%%%%%%%%%%%%%%%%%%%%%%%%%%%%%%%%%%%%%%%%%%%%%%%%%%%%%%%%%%%%
\section{Brownian dynamics simulations}
\label{app:bdsim}

Brownian dynamics simulations are used to verify our simplified model for dynamics over a free energy landscape in the $\rho$ and $r$ coordinates. We define a discretized version of the heterogeneous worm-like chain model, using the standard bead-rod formalism \cite{somasi2002brownian}, with very stiff stretching modulus for constraining the length of the rods. Our chains consist of $n=20$ segments of length $d$, with bending energy 
\begin{equation}
E_\text{bend}=\sum_{i=1}^{n-1} \kappa_i\left[1-\cos(\rho_i)\right]
\end{equation}
for $\rho_i = \cos\theta_i$ and $\theta_i$ the angle between orientations of each consecutive pair of segments. The prefactor is set to $\kappa_i = \frac{\ell_p,1}{d}$ for $i\leq 10$ and $\kappa_i = \frac{\ell_p,2}{d}$ otherwise. The central bead represents a junction of size $\Delta = d$.

Chains are initiated in a thermally equilibrated configuration by direct sampling of the segment angles. A standard Brownian dynamics algorithm \cite{ermak1978brownian} with 4th-order Runge-Kutta time integration\cite{koslover2014multiscale} is used to propagate the system forward in timesteps of $\delta t = 10^{-4} \frac{d^2 \mu_b }{k_BT}$, where the $\mu_B$ is the friction coefficient of each bead. Simulations are run until either the central chain angle $\rho_{10}$ or the end-to-end distance reaches a cutoff value, up to a maximum of $10^7$ timesteps. 
%by sampling each $\rho$ from the proper distribution, and then picking a random direction to point in. For every time step forward, we determine the new position of The force exerted on each individual bead comes from a combination of a random Brownian force and the restorative force of the chain trying to straighten out. The bead positions are updated every time step using fourth order runge-kutta method. We repeat this procedure on the chain until it reaches a conformation where one of the $n$ beads reaches some cutoff energy and breaks. We average these results to obtain our mean time to break.

Mean first passage times to cutoff cannot be obtained by direct averaging since many chains to not reach the cutoff over the simulation time. Instead, we fit the empirical cumulative distribution function for first passage times to the functional form $1 - \exp(-t/\tau)$, to extract the appropriate time-scale for first passage. $10^4$ chains are simulated for each data point plotted in Fig.~\ref{fig:bdsim}.

%%%%%%%%%%%%%%%%%%%%%%%%%%%%%%%%%%%%%%%%%%%%%%%%%%%%%%%%%%%%%%%%%%%%%
\bibliographystyle{aip} 
\bibliography{filamentfpt} 
%%%%%%%%%%%%%%%%%%%%%%%%%%%%%%%%%%%%%%%%%%%%%%%%%%%%%%%%%%%%%%%%%%%%%%

\end{document}